\documentclass[preprintnumbers,nofootinbib,noshowpacs,eqsecnum,prd,superscriptaddress,letterpaper]{revtex4}

\usepackage{graphicx}
\usepackage{amsmath,amssymb}
\usepackage{url}
\usepackage{multirow}
\usepackage{hyperref,color}
\usepackage[normalem]{ulem}
\graphicspath{{figs/}}
\usepackage{xcolor}
\usepackage{slashed}
\usepackage{rotating}

\newcommand{\sla}[1]{/\!\!\!\!#1}

\preprint{\bf YITP-SB-2021-25}

\newcommand{\UH}{\mathbf{U}}
\newcommand{\nq}{\mathcal{N}^\mathcal{Q}}
\newcommand{\nl}{\mathcal{N}^\mathcal{\ell}}
\newcommand{\rqq}{\mathcal{R}^\mathcal{Q}}
\newcommand{\rl}{\mathcal{R}^\mathcal{\ell}}
\newcommand{\rql}{\mathcal{R}^\mathcal{Q\ell}}

\def\cF{{\cal F}}

\def\cQ{{\cal Q}}
\def\cP{{\cal P}}

\def\cY{{\cal Y}}
\def\Tr{{\rm Tr}}

\newcommand{\DLR}{\mathbf{D}}

\newcommand{\g}{\gamma}

\newcommand{\de}{\partial}
\newcommand{\hc}{\text{h.c.}}

\newcommand{\tr}{\Tr}

\DeclareMathOperator{\diag}{diag}
\newcommand{\U}{\mathbf{U}}
\newcommand{\T}{\mathbf{T}}
\newcommand{\V}{\mathbf{V}}

\newcommand{\WWd}{W_{\mu\nu}}
\newcommand{\WWu}{W^{\mu\nu}}
\newcommand{\BBd}{B_{\mu\nu}}
\newcommand{\BBu}{B^{\mu\nu}}
\newcommand{\GGd}{\mathcal{G}_{\mu\nu}}
\newcommand{\GGu}{\mathcal{G}^{\mu\nu}}
\newcommand{\QBL}{\bar{Q}_L\,}
\newcommand{\QBR}{\bar{Q}_R\,}
\newcommand{\QL}{\,Q_L\,}
\newcommand{\QR}{\,Q_R\,}
\newcommand{\LBL}{\bar{L}_L\,}
\newcommand{\LBR}{\bar{L}_R\,}
\newcommand{\LL}{\,L_L\,}
\newcommand{\LR}{\,L_R\,}

\def\beq{\begin{equation}}
\def\eeq{\end{equation}}

\newcounter{twofermionQ}

\newcounter{twofermionL}

\newcounter{fourfermionQ}

\newcounter{fourfermionL}

\newcounter{fourfermionQL}

\begin{document}
\title{Electroweak HEFT after LHC run 2}
\author{Oscar J. P. \'Eboli}
\email{eboli@if.usp.br}
\affiliation{Instituto de F\'isica, Universidade de S\~ao Paulo,
  R. do Mat\~ao 1371, 05508-090 S\~ao Paulo, Brazil}
\author{M.~C.~Gonzalez--Garcia}
\email{maria.gonzalez-garcia@stonybrook.edu}
\affiliation{Departament de Fisica Quantica i Astrofisica and Institut
  de Ciencies del Cosmos, Universitat de Barcelona, Diagonal 647,
  E-08028 Barcelona, Spain}
\affiliation{Instituci\'o Catalana de Recerca i Estudis
  Avancats (ICREA) Passeig Llu\'is Companys 23, 08010 Barcelona,
  Spain.}  \affiliation{C.N. Yang Institute for Theoretical Physics,
  Stony Brook University, Stony Brook New York 11794-3849, USA}
\author{Matheus Martines}
\email{matheus27martines@gmail.com}
\affiliation{Instituto de F\'isica, Universidade de S\~ao Paulo,
  R. do Mat\~ao 1371, 05508-090 S\~ao Paulo, Brazil}
\affiliation{Departament de Fisica Quantica i Astrofisica and Institut
  de Ciencies del Cosmos, Universitat de Barcelona, Diagonal 647,
  E-08028 Barcelona, Spain}

\begin{abstract}

  We analyze the electroweak interactions in the framework of the
  Higgs effective field theory using the available Higgs and
  electroweak diboson production results from LHC run 2 as well
  as the electroweak precision data. Assuming universality of the weak
  current, our study considers 25 possible anomalous couplings. To
  unveil the nature of the Higgs boson, {\em i.e.} isosinglet versus
  part of $SU(2)_L$ doublet, we explore the correlation effects
  between observables that are predicted to exist in the linear
  realization of the electroweak gauge symmetry but not in its
  non-linear counterpart. This improves previous studies aimed at
  investigating the Higgs nature and the origin of the electroweak
  symmetry breaking.

\end{abstract}

\maketitle

\section{Introduction}
\label{sec:intro}

The CERN Large Hadron Collider (LHC) has accumulated an impressive
amount of data which allows not only to search for direct beyond the
standard model (BSM) signals, but also to perform precision tests of
the standard model (SM). Due to the lack of direct evidence of new
physics, it is reasonable to assume that it is heavy. Therefore, we
use effective Lagrangians~\cite{Weinberg:1978kz, Georgi:1985kw,
  Donoghue:1992dd} to search for deviations from the SM
predictions. \medskip

The nature of the Higgs-like state observed at the LHC in
2012~\cite{Aad:2012tfa, Chatrchyan:2012xdj} plays a pivotal role in
the construction of the low-energy effective field theory. If it
belongs to a $SU(2)_L$ doublet, the SM gauge symmetry can be realized
linearly in the effective theory, which in this case is called
standard model effective field theory (SMEFT). Conversely, if the
Higgs boson is a $SU(2)_L$ isosinglet, we are led to use a non-linear
realization of the gauge symmetry and the low effective theory
obtained this way is called Higgs effective field theory (HEFT); for a
review of these frameworks see Ref.~\cite{Brivio:2017vri}.  In a
top-down approach the Wilson coefficients depend upon the specific
UV-completion realized in nature. Here, we adopt a bottom-up strategy
where the Wilson coefficients are treated as free parameters,
therefore, probing a large set of models simultaneously.  \medskip

In this work we analyze the presently available electroweak data to
study the HEFT. More specifically, we employ the $Z$ pole precision
electroweak observables (EWPO), the diboson $WW$, $\gamma W$ and $WZ$
productions and the recently released Higgs kinematic distributions to
constrain the HEFT Wilson coefficients.  Furthermore, we also study
the Higgs nature by analyzing the correlations between the diboson
production and Higgs properties that are present in the linear
realization of the symmetry but they are absent in the non-linear
one~\cite{Brivio:2013pma}.  We also study the impact of the LHC data
on a class of composite Higgs models~\cite{Kaplan:1983fs,
  Kaplan:1983sm, Agashe:2004rs} which introduce correlations between
some of the HEFT Wilson coefficients.\medskip

Previous works analyzed the LHC HEFT phenomenology taking into account
a different choice of power counting to order the HEFT
series~\cite{Buchalla:2013eza}; see for instance
Refs.~\cite{Gillioz:2012se, Ellis:2013lra, Azatov:2012bz,
  Buchalla:2015qju, Corbett:2015ksa, Li:2019ghf, Liu:2019rce,
  deBlas:2018tjm}.  Generically, these works consider anomalous
interactions which exhibit the same Lorentz structure of the SM.
Here, we expand the list of effective operators probed in these
analyses and also consider substantially updated datasets with most of
the experimental results containing the full LHC run 2 luminosity. In
particular we include in the analysis the most up-to-date results on
the kinematic distributions for the Higgs observables in the form of
simplified template cross sections
(STXS)~\cite{LHCHiggsCrossSectionWorkingGroup:2016ypw,Andersen:2016qtm}.
Furthermore we not only consider constraints associated with Higgs
observables but we also analyze the impact of the LHC run 2 data on
TGCs. This allows us quantify observables which depend on the nature
of the Higgs boson. Finally, our quantification of the present status
of the bounds on some specific composite Higgs models extends that of
previous works~\cite{Banerjee:2017wmg, Sanz:2017tco, deBlas:2018tjm,
  Khosa:2021wsu} by considering the impact of the Higgs kinematic
distributions and the full LHC run 2 dataset. \medskip

The overall picture that emerges from our analyses is that the
presently available data is in agreement with the SM, as could be
anticipated. This allows us to obtain stringent constraints on the
Wilson coefficients parametrizing our bottom-up approach. In
particular, the Higgs interactions to photon and gluon pairs receive
the strongest constraints. In addition the triple electroweak gauge
boson couplings agree with the SM prediction at the percent
level. \medskip

This work is organized as follows. We present in
Sec.~\ref{sec:thframe} the theoretical framework of our studies while
Sec.~\ref{sec:frame} describes how we performed our analyses as well
as the datasets used in it.  Our results are presented in
Sec.~\ref{sec:results} and we discuss some implication of those in
Sec.~\ref{sec:disc}. \medskip

\section{Theoretical Framework}
\label{sec:thframe}

In this work we consider that the observed Higgs-like state is an
isosinglet of the SM gauge symmetries. In this scenario, the
realization of the SM gauge symmetries is non-linear. Denoting the EW
Goldstone bosons (GB) as $\vec{\pi}$, the building block of the
low-energy effective Lagrangian is a dimensionless unitary matrix
\begin{equation}
\UH(x) = e^{i \sigma^a \pi_a(x)/f} 
\label{eq:def:u}
\end{equation}
where $f$ is the GB scale and $\sigma^a$ are the Pauli matrices.
$\UH$ transforms as a bidoublet under the global symmetry
$SU(2)_L \otimes SU(2)_R$
\begin{equation}
\UH \to \UH^\prime = L U R^\dagger
\end{equation}
with $L$ ($R$) being a $SU(2)_{L(R)}$ transformation. Its covariant
derivative reads
\begin{equation}
\DLR_\mu \UH(x) \equiv \partial_\mu \UH(x) +ig \frac{\sigma^j}{2}
W^j_{\mu}(x)\UH(x) - \frac{ig'}{2}  B_\mu(x) \UH(x)\sigma_3 \, .
\end{equation}
It is also convenient to define the vector chiral field and its
covariant derivative as
\begin{eqnarray}
\V_\mu&\equiv&  \left(\DLR_\mu\UH\right)\UH^\dagger \,,
  \\
  && \nonumber\\
D_\mu \V_\alpha &=& \partial_\mu \V_\alpha + i g [ W_\mu , \V_\alpha]  \,,
\end{eqnarray}
where $W_\mu$ stands for $W^j_\mu \,\sigma^j/2$.  We also define the
scalar chiral field $\T\equiv\UH\sigma_3\UH^\dag$. These three objects
transform in the adjoint of $SU(2)_L$. \medskip

We follow the notation defined in Ref.~\cite{Brivio:2016fzo} to which
we refer the reader for details in the construction of the HEFT
Lagrangian.  In brief, the leading order (LO) term in the HEFT
expansion is
\begin{equation}
\begin{split}
{\cal L}_0=& -\dfrac{1}{4} \GGd^\alpha \mathcal{G}^{\alpha\,\mu\nu}
-\dfrac{1}{4} \WWd^a W^{a\,\mu\nu}-\dfrac{1}{4} \BBd\BBu+\\
&+\dfrac{1}{2}\de_\mu h \de^\mu h-\dfrac{v^2}{4}\Tr(\V_\mu \V^\mu)\cF_C(h)-V(h)
+\\
&+i\bar{Q}_L\slashed{D}Q_L+i\bar{Q}_R\slashed{D}Q_R+i\bar{L}_L\slashed{D}L_L
+i\bar{L}_R\slashed{D}L_R+\\
&-\dfrac{v}{\sqrt2}\left(\bar{Q}_L\U \cY_Q(h) Q_R+\hc\right)
-\dfrac{v}{\sqrt2}\left(\bar{L}_L\U \cY_L(h) L_R+\hc\right) \,,
\end{split}
\label{eq:lag0}
\end{equation}
where $L$ ($Q$) denotes the lepton (quark) fermionic field.
$\GGd^\alpha$, $\WWd^a$ and $\BBd$ stand for the $SU(3)_c$, $SU(2)_L$
and $U(1)_Y$ field strengths respectively.  The function $\cF_C(h)$
can be expanded as
\begin{equation}\label{FC}
 \cF_C(h) = 1 + 2\, a_C\, \frac{h}{v}+b_C \,\frac{h^2}{v^2}+\dots
\end{equation} 
where the dots account for higher powers of $(h/v)$. It is convenient
to single out the BSM part of the coefficients $a_C$ and $b_C$ by
writing
\begin{equation}
 a_C = 1 + \Delta a_C\,,\qquad 
 b_C = 1 + \Delta b_C\,,
\end{equation} 
where $\Delta a_C,\, \Delta b_C$ is assumed to be of the same order as
the coefficients accompanying the operators appearing in
next-to-leading order (NLO) $\Delta {\cal L}$.  \medskip

The Yukawa couplings depend on functions $\cY_{Q,L}(h)$ analogous to
$\cF_C(h)$, whose first two terms are:
\begin{equation}
\begin{aligned}
  \cY_{Q}(h)\equiv& \diag\left( Y_U^{(0)} + 
 Y_{U}^{(1)}\dfrac{h}{v} + \dots \;,\;
Y_{D}^{(0)} +Y_{D}^{(1)}\dfrac{h^n}{v^n} + \dots\right)\,,\qquad
\cY_{L}(h)\equiv& \diag\left( 0 \;,\;  Y_{\ell}^{(0)} +  Y_{\ell}^{(1)}\dfrac{h}{v}+\dots\right)\,.
\end{aligned}
\end{equation}
The $Y^{(0)}$ terms yield fermion masses, while the $Y^{(1)}$ ones
control the Higgs coupling to fermion pairs. \medskip

The BSM contributions are described by a $\Delta {\cal L}$ whose
ordering depends upon the choice of power counting. Following
Ref.~\cite{Brivio:2016fzo}, we consider that the NLO operators are
either the ones necessary to renormalize one-loop divergences or
receive finite one-loop contributions. In total, in the absence of
right-handed neutrinos there are 148 independent operators that
conserve $CP$ without taking into account flavor indices.\medskip

This Lagrangian can be generically written as
\begin{equation}
  \Delta {\cal L}= \sum_{i} \,c_i \,\cP_i\,+\,
  \sum_{i}\, n^{\cal Q}_i\, \nq_i \,+\,
  \sum_{i} \,n^{\cal \ell}_i \,\nl_i \,+\,
  \sum_{i}\, r^{\cal Q}_i \,\rqq_i \,+\,
  \sum_{i} r^{\cal \ell}_i \rl_i +
  \sum_{i} r^{\cal Q\ell}_i \rql_i   
\end{equation}
for operators involving only bosons ($\cP_i\,$), two quark or two
lepton currents ($\nq$, and $\nl$) and four fermion currents ($\rqq$,
$\rl$ and $\rql$).  Each of them includes a function $\cF_i(h)$
conventionally parametrized as
\begin{equation}
 \cF_i(h) =1 + 2\,\bar{a}_i\,\frac{h}{v}+\bar{b}_i\,\frac{h^2}{v^2}+\dots
\end{equation}
If $c_j$ is the Wilson coefficient of the effective operator $j$, it
appears multiplying all the terms in the definition of
$\cF_j$. Therefore, for convenience, in what follows we introduce the
notation
$$c_j \,\bar a_j \to a_j\, , \;\;\;\; c_j \,\bar b_j \to b_j\;\;\dots$$ 
and equivalently for operators with coefficients $n_j$ and
$r_j$. \medskip

At this point, it is important to notice that there are not enough
data to constrain all the Wilson coefficients contained in the LO and
NLO Lagrangians, therefore we focus on a representative subset of
operators.  From the LO Lagrangian we shall consider 4 of the Yukawa
couplings together with the bosonic operator $\cF_C$. At NLO we
include effects from 13 purely bosonic operators and 7 operators
involving fermions.\medskip

In particular we consider the 10 operators that contribute to the
electroweak precision data
\begin{equation}
  \begin{aligned}
    \cP_{1}(h) &= \BBd \tr(\T \WWu) \cF_{1} \,,
\\
    \cP_T(h) &= \frac{v^2}{4} \tr(\T\V_\mu)\tr(\T\V^\mu) \cF_{T}\,,
\\
    \cP_{12}(h) &= (\tr(\T\WWd))^2 \cF_{12} \,,
\\
\nq_1&\equiv i\QBL \g_\mu \V^\mu \QL \cF_{1Q}	 \,,	
\\
\nq_2 + \nq_8 &= i\QBR \g_\mu \U^\dag\V^\mu\U \QR \cF_{2Q} 
+ i\QBR \g_\mu \U^\dag\T \V^\mu\T \U\QR \cF_{8Q} \,,
\\
\nq_5 & =  i\QBL \g_\mu \{\V^\mu,\T\}\QL \cF_{5Q} \,,
\\
\nq_6 &=  i\QBR \g_\mu \U^\dag\{\V^\mu,\T\}\U\QR \cF_{6Q} \,,
\\
\nq_7 & = i\QBL \g_\mu \T\V^\mu\T\QL \cF \,,
\\
\nl_2 & =  i\LBR \g_\mu \U^\dag\{\V^\mu,\T\}\U\LR \cF_{2\ell} \,,
\\
\rl_2 - \rl_5 & = (\LBL\g_\mu\LL)(\LBL\g^\mu\LL) \cF_{2L}
- (\LBL\g_\mu\T\LL)(\LBL\g^\mu\T\LL) \cF_{5L} \,.
\end{aligned}
\label{eq:oper-ewpd}
\end{equation}

The operators $\cP_1$, $\cP_T$, $\cP_{12}$, and $\rl_2 - \rl_5 $
contribute to the oblique parameters S, T, U, and to a shift to the
Fermi constant respectively. Moreover, the six remaining operators
modify the $W$ and $Z$ couplings to fermion pairs. These contributions
are presented in detail in Ref.~\cite{Brivio:2016fzo}. \medskip

Altogether in our fit to the EWPO we parametrize the HEFT
contributions in terms of the following Wilson coefficients
\begin{equation}
  \left \{
    c_1 \;,\; c_T \;,\; c_{12} \;,\; 
    n_1^\cQ \;,\; n_2^\cQ  + n^\cQ_8 \;,\; n_5^\cQ \;,\; n_6^\cQ \;,\; n_7^\cQ \;,\; n_2^\ell \;,\;
    r^\ell_2-r^\ell_5
  \right\} \;,
\label{eq:ewpdc}
\end{equation}
which correspond to the contributing part of the operators in
Eq.~\eqref{eq:oper-ewpd} by taking $\cF=1$. \medskip

In addition, we consider 4 bosonic operators that modify the triple
electroweak gauge boson couplings (TGC) and affect the production of
electroweak gauge boson pairs $W^+ W^-$, $ W^\pm Z$ and $W^\pm \gamma$
at LHC:
\begin{equation}
\begin{aligned}
  &\cP_{2}(h) = \dfrac{i}{4\pi}\BBd \tr(\T[\V^\mu,\V^\nu]) \cF_{2} \;,
\\
&\cP_{3}(h)= \dfrac{i}{4\pi}\tr(\WWd [\V^\mu,\V^\nu]) \cF_{3} \;,
\\
&\cP_{13}(h) = \dfrac{i}{4\pi} \tr(\T\WWd)\tr(\T[\V^\mu,\V^\nu])\cF_{13} \;,
\\
&\cP_{WWW}(h)
=\dfrac{4\pi\varepsilon_{abc}}{\Lambda^2}W_\mu^{a\nu}W_\nu^{b\rho}
W_{\rho}^{c\mu}\cF_{WWW} \;,
\end{aligned}
\end{equation}
whose correspondent Wilson coefficients for $\cF=1$ are
\begin{equation}
  \left \{
    c_2 \;,\; c_3 \;,\; c_ {13} \;,\, c_{WWW}
\right \} \;.
\label{eq:tgvc}
\end{equation}
These operators modify the triple gauge couplings (TGC)
$\gamma W^+W^-$ and $ZW^+W^-$.  These anomalous contributions can be
generically parametrized in terms of the the usual effective TGC
Lagrangian given in Ref.~\cite{Hagiwara:1986vm}:
\begin{equation}
{\cal L}_{WWV} = - \,i g_{WWV} \Bigg\{ 
g_1^V \Big( W^+_{\mu\nu} W^{- \, \mu} V^{\nu} - 
W^+_{\mu} V_{\nu} W^{- \, \mu\nu} \Big) 
   \,+\, \kappa_V W_\mu^+ W_\nu^- V^{\mu\nu}
+\frac{\lambda_V}{2 m_W^2} \; W_{\mu \nu}^+ W^{- \nu \rho} V_\rho^\mu
\Bigg\}\,, \label{eq:tgveff} 
\end{equation}
with deviations from the SM predictions
$g_1^Z=\kappa_Z=\kappa_\gamma=1$, $\lambda_\gamma=\lambda_Z=0$
\begin{eqnarray}
 && \Delta g_1^Z =g_1^Z-1\equiv\frac{g}{4\pi c_W^2} c_3\,, 
\nonumber \\
&&  \Delta \kappa_Z= \kappa_Z-1\equiv\frac{g}{4\pi} \left(c_3 +2
   c_{13} - 2t_W  c_2\right)\,, \label{eq:kappa}
\\
&&  \Delta \kappa_\gamma  =\kappa_\gamma-1\equiv\frac{g}{4\pi} 
\left(c_3 +2 c_{13}+ 2 
\frac{c_2}{t_W}\right)\, , 
\nonumber \\ 
&&\lambda_\gamma=\lambda_Z\equiv\frac{6 \pi\, g\, v^2}{\Lambda^2} c_{WWW}
\,, \nonumber
\end{eqnarray}
where $c_W$ ($t_W$) stands for $\cos\theta_W$ ($\tan\theta_W$).
Electromagnetic gauge invariance enforces $g_1^\gamma=1$, both in the
SM and in the presence of the new operators.  In
Eq.~\eqref{eq:tgveff}, $V \equiv \{\gamma, Z\}$, $g_{WW\gamma} = e$,
$g_{WWZ} = g \cos\theta_W$, and $W^\pm_{\mu\nu}$ and $V_{\mu\nu}$
refer exclusively to the kinetic part of the gauge field
strengths. \medskip

Concerning Higgs processes, eleven additional operators take part in
our Higgs couplings analysis. They originate from the part
proportional to $\Delta a_C$ of the operator $\cF_C$ and the
deviations of the Yukawa couplings ($Y^{(1)}_f$) for the top, bottom,
tau, and muon in Eq.~\eqref{eq:lag0} as well as the NLO bosonic
operators
\begin{equation}
\begin{aligned}
  &\cP_{4}(h) =\dfrac{i}{4\pi} \BBd \tr(\T\V^\mu) \de^\nu\cF_{4} 
\\
&\cP_{5}(h)=\dfrac{i}{4\pi} \tr(\WWd\V^\mu) \de^\nu\cF_{5} \,,
\\
&\cP_{17}(h) =\dfrac{i}{4\pi} \tr(\T \WWd) \tr(\T\V^\mu)
\de^\nu\cF_{17} \,,
\\
& \cP_{B}(h) =-\dfrac{1}{4}\BBd \BBu \cF_B \,,
\\
&\cP_{W}(h) =-\dfrac{1}{4}\WWd^a W^{a\mu\nu} \cF_W \,,
\\
&\cP_G(h) = -\dfrac{1}{4}G_{\mu\nu}^a G^{a\mu\nu}\cF_G\,.
\end{aligned}  
\end{equation}
The corresponding Wilson coefficients characterizing the strength of
the effective interaction that enter in the Higgs analysis are
\begin{equation}
\left \{ a_4 \;,\; a_5 \;,\; a_{17} \;,\;
  a_B \;,\; a_W \;,\; a_G
\right\} \,.
\label{eq:higgc}
\end{equation}
We notice that, in principle, $a_{T (1)}$ also affects the Higgs
couplings.  However, we anticipate here that the EWPO will set strong
bounds on $c_{T (1)}$, and consequently, under the assumption that the
corresponding $\bar a_{T(1)}$ couplings are at most ${\cal O}(1)$, we
can safely neglect their contribution. \medskip Notice also that, for
the sake of simplicity, we have not consider the deviation on Higgs
quartic vertices ($HVf\bar f'$) generated by some of the fermion
current operators. \medskip

As we will see in the following section, at present there is enough
experimental information to individually bound the 20 Wilson
coefficients in Eqs.~\eqref{eq:ewpdc},~\eqref{eq:tgvc},
and~\eqref{eq:higgc} as well as 4 Yukawa couplings and $\Delta a_C$.
However, the analysis, in particular when performed up to quadratic
order in the coefficients, can potentially exhibit discrete (quasi-)
degeneracies associated to sign flips of the SM Higgs couplings. For
example, the vertex $HV^\mu V_\mu$ ($V=W^\pm$ or $Z$) is proportional
to~\cite{Brivio:2016fzo}
\begin{equation}
  1 + \Delta a_C   \;
\label{eq:vvdeg}
\end{equation}
therefore, we can anticipate a degeneracy with the SM results for the
$HV^\mu V_\mu$ vertex ($\Delta a_C=0$) for $\Delta a_C=-2$. \medskip

In similar fashion, the anomalous interactions can also lead to Yukawa
couplings of the order of the SM ones but with a different sign
because the coefficient of the $H \bar{f} f$ vertex is now
\begin{equation}
  \frac{1}{\sqrt{2}}Y_f^{(1)}=\frac{1}{\sqrt{2}}Y_f^{(0)}\;\left(
  \frac{Y_f^{(1)}}{Y_f^{(0)}}\right)\; ,
\label{eq:yukdeg}  
\end{equation}
and, therefore, it presents one degenerate solution
$\frac{Y_f^{(1)}}{Y_f^{(0)}}=- 1$ with the SM one
$\frac{Y_f^{(1)}}{Y_f^{(0)}}=+1$. \medskip

Another source of degeneracy is the effective photon-photon-Higgs
coupling that gets corrections from by $\cP_W$ and $\cP_B$:
\begin{equation}
  -\frac{1}{4} G_{SM}^{\gamma\gamma} + \frac{1}{2v} (a_B c_W^2 + a_W s_W^2)
  \label{eq:gagadeg}  
\end{equation}
where $G_{SM}^{\gamma\gamma} \simeq 3.3 \times 10^{-2}$ is the
one-loop SM contribution.  Consequently, a SM-like solution for the
Higgs decay into $\gamma\gamma$ can be found for
$ a_B c_W^2 + a_W s_W^2 \simeq v\, G_{SM}^{\gamma\gamma} \simeq 8.4
\times 10^{-3}$. \medskip

A similar effect is also present in  $H\GGu\GGd$ whose coupling
in the large top mass limit is 
\begin{equation}
  - \frac{1}{4} G_{SM}^{gg} \left(\frac{Y_t^{(1)}}{Y_t^{(0)}}\right)-\frac{1}{2v}\; a_G\;. 
  \label{eq:glugludeg}  
\end{equation} 
with $G_{SM}^{gg} \simeq 5.3 \times 10^{-2}$ TeV$^{-1}$ being the
one-loop SM contribution. \medskip

It is important to notice that the effective operators entering the
degeneracies in Eqs.~\eqref{eq:vvdeg}--\eqref{eq:glugludeg} lead to
distinct Higgs kinematic distributions either due to a different
number of derivatives of the Higgs field or their contribution to the
one-loop Higgs coupling to gluon pairs~\cite{Buschmann:2014twa}.
Therefore, we anticipate that some of the potential degeneracies can
be resolved by the available data on the Higgs kinematic
distributions, as we will see in Sec.~\ref{sec:results}. \medskip


As it is well known, when the electroweak gauge symmetry is linearly
realized in the low-energy effective theory, the Higgs boson is part
of a $SU(2)_L$ doublet as in the SM. This is the scenario described by
the SMEFT. The comparison between the results of the analyses
performed in the frameworks of HEFT and SMEFT allows us to probe the
nature of the Higgs boson~\cite{Brivio:2013pma}. A particularly
sensitive probe is associated with the (de)correlation of the
contributions to the TGC and Higgs-gauge-boson couplings in the two
scenarios. In brief, in SMEFT the following operators induce anomalous
triple gauge boson couplings
\begin{equation}
\begin{aligned}
&{\cal O}_{W}=\dfrac{ig}{2}(D_{\mu} \Phi)^{\dagger} W^{\mu
  \nu} (D_{\nu} \Phi)\,,\qquad\qquad
\\
&{\cal O}_{B}=\dfrac{ig'}{2}(D_{\mu} \Phi)^{\dagger}
B^{\mu \nu} (D_{\nu} \Phi)\,,
\end{aligned}
\end{equation}
and, in this framework, the linear realization of the gauge symmetry
implies that the same operators give rise to correlated corrections to
the Higgs couplings to the electroweak vector bosons.  On the
contrary, in HEFT there are four sibling operators, two of them,
$\cP_2$ and $\cP_3$, giving the corresponding corrections to the TGC
(proportional to $c_2$ and $c_3$) and the other two, $\cP_4$ and
$\cP_5$, inducing the corresponding shift for the $HVV$ vertex
(proportional to $a_4$ and $a_5$). It is clear then that in HEFT there
is no {\em a priori} connection between the corrections to the TGC and
the $HVV$ vertices. \medskip

Following Refs.~\cite{Brivio:2013pma,Brivio:2016fzo} we construct four
specific combinations of the coefficients $c_2$, $c_3$, $a_4$, and
$a_4$ which are useful for quantifying the status of these
(de)correlations in the Higgs and TGC results
\begin{equation}
\begin{aligned}
&\Sigma_B\equiv \frac{1}{\pi g t_W}(2c_2+a_4)\,, 
\qquad\qquad 
\Sigma_W\equiv \frac{1}{2\pi g}(2c_3-a_5)\,,\\
&\Delta_B\equiv \frac{1}{\pi g t_W}(2c_2-a_4)\,, 
\qquad\qquad 
\Delta_W\equiv \frac{1}{2\pi g}(2c_3+a_5)\,.
\label{eq:sigdel}
\end{aligned}
\end{equation}
These four parameters were defined in such a way that, at
dimension-six order in the SMEFT expansion, the two $\Delta$'s are
zero because of gauge invariance and of the doublet nature of the
Higgs, $\Delta_B=\Delta_W=0$. Moreover, the $\Sigma$'s are directly
related to the Wilson coefficients of the operators ${\cal O}_{W}$ and
${\cal O}_{B}$: $\Sigma_B=v^2\frac{f_B}{\Lambda^2}$ and
$\Sigma_W=v^2\frac{f_W}{\Lambda^2}$, being $f_i$ the associated Wilson
coefficient. In contrast, the HEFT operators can generate independent
modifications to each of these four variables. Therefore, the study of
these parameters can shed light on the nature of the Higgs
boson. \medskip


In a top-down approach, we also analyze the minimal composite Higgs
scenario~\cite{Agashe:2004rs, Carena:2014ria} that is based on global
symmetry $SO(5)$ broken to $SO(4)$ at scale $f$.  In this model the
Higgs interaction to vector bosons is modifed with respect to the SM
by a multiplictive factor
\begin{equation}
  a_C = \sqrt{1 - \xi}
  \label{eq:amod}
\end{equation}
with $\xi = v^2/f^2$. The modification to the Higgs couplings to
fermions
\begin{equation}
c_p  \equiv \frac{Y_p^{(1)}}{Y_p^{(0)}} 
\end{equation}  
depends on how the SM fermions are embedded in the UV-theory.  Here,
following~\cite{Sanz:2017tco}, we consider the two characteristic
choices labeled $A$ and $B$ :
\begin{equation}
  c_p^A  = \sqrt{1-\xi} \;\;\;\hbox{or }\;\;\;
  c_p^B = \frac{1-2\,\xi}{\sqrt{1-\xi}} \,,
\label{eq:cmod}
\end{equation}  
where in the equations above $p$ stands for top, bottom, tau and muon
in our analysis.  \medskip

\section{Analysis Framework}
\label{sec:frame}


In order to obtain the present constraints on HEFT, we considered the
available data on the EWPO, the triple electroweak gauge couplings and
the Higgs data.  In the EWPO data analysis, we analyze 15 observables
of which 12 are $Z$ observables~\cite{ALEPH:2005ab}:
\begin{equation}
\Gamma_Z \;\;,\;\;
\sigma_{h}^{0} \;\;,\;\;
{\cal A}_{\ell}(\tau^{\rm pol}) \;\;,\;\;
R^0_\ell \;\;,\;\;
{\cal A}_{\ell}({\rm SLD}) \;\;,\;\;
A_{\rm FB}^{0,l} \;\;,\;\;
R^0_c \;\;,\;\;
 R^0_b \;\;,\;\;
{\cal  A}_{c} \;\;,\;\;
 {\cal A}_{b} \;\;,\;\;
A_{\rm FB}^{0,c}\;\;,\;\;
\hbox{ and} \;\;
A_{\rm FB}^{0,b}  \hbox{ (SLD/LEP-I)}\;\;\; ,
\end{equation}
supplemented by three $W$ observables: the $W$ mass ($M_W$) taken
from~\cite{Olive:2016xmw}, its width ($\Gamma_W$) from
LEP2/Tevatron~\cite{ALEPH:2010aa} and the leptonic $W$ branching ratio
($\hbox{Br}( W\to {\ell\nu})$)~\cite{Olive:2016xmw}. In our
statistical analysis we define the chi-square function
\begin{equation}
  \chi^2_{\rm EWPO} (    c_1 \;,\; c_T \;,\; c_{12} \;,\; 
    n_1^\cQ \;,\; n_2^\cQ  + n^\cQ_8 \;,\; n_5^\cQ \;,\; n_6^\cQ \;,\; n_7^\cQ \;,\; n_2^\ell \;,\;
    r^\ell_2-r^\ell_5
)
\end{equation}  
and fitted the relevant Wilson coefficients. Notice that we assumed
the couplings to be generation independent and diagonal in flavor
space; for further details on this analysis see
Ref.~\cite{Brivio:2016fzo}. \medskip


In order to study the triple couplings of electroweak gauge bosons we
considered electroweak diboson data (EWDBD) from LHC, more
specifically the diboson production of $WZ$, $WW$ and $W\gamma$ pairs
as well as the vector boson fusion production of $Z$'s ($Zjj$). The
specific data employed in our study is presented in the top rows of
Table~\ref{tab:tgv-h-data}. In total we considered 73 data points in
this analysis. \medskip

The theoretical predictions were obtained by simulating the $W^+W^-$,
$W^\pm Z$, $W^\pm \gamma$, and $Zjj$ channels that receive
contributions from TGC.  To this end, we used
\textsc{MadGraph5\_aMC@NLO}~\cite{Frederix:2018nkq} with the UFO files
for our effective Lagrangian generated with
\textsc{FeynRules}~\cite{Christensen:2008py, Alloul:2013bka}.  We
employ \textsc{PYTHIA8}~\cite{Sjostrand:2007gs} to perform the parton
shower and hadronization, while the fast detector simulation is
carried out with \textsc{Delphes}~\cite{deFavereau:2013fsa}.  Jet
analyses were performed using
\textsc{FASTJET}~\cite{Cacciari:2011ma}.\medskip

These predictions are statistically confronted with the LHC run 2 data
by constructing a binned log-likelihood function based on the data
contents of the different bins in the kinematic distribution of each
channel. We consistently take into account not only the statistical
errors but also the systematic and theoretical uncertainties adding
them in quadrature and assuming some partial correlation among them
which we estimate with the information provided by the
experiments. \medskip

For the sake of simplicity in our EWDBD analysis we discarded possible
effects from anomalous couplings of gauge bosons to fermion pairs,
which are well constrained by the EWPO data so we have
\begin{equation}
\chi^2_{\rm TGC} (c_2 \;,\; c_3 \;,\; c_ {13} \;,\, c_{WWW}) \;.
\end{equation}
%

\begin{table} [ht]
\begin{tabular}{|@{\hskip 0.5cm}c|l|l|c|l|l|}
\hline 
& Channel ($a$) & Distribution & \# bins   &\hspace*{0.2cm} Data set & \hspace*{0.2cm}Int Lum  \\ [0mm]
  \hline
  & $WZ \to \ell^+ \ell^- \ell^{\prime\pm}$ & {$M(WZ)$} & 7& CMS 13
       TeV,  &  137.2 fb$^{-1}$~\cite{CMS:2021lix} \\[0mm]
  \multirow{7}{*}
{\begin{rotate}{90}   EWDBD\end{rotate}}
&$WW \to \ell^+\ell^{(\prime)-}+ 0/1 j$   &$M(\ell^+\ell^{(\prime)-})$
                             &11 & CMS 13 TeV, & 35.9 fb$^{-1}$~\cite{CMS:2020mxy} \\[0mm]
&  $W\gamma \to \ell \nu \gamma$ & $\frac{d^2\sigma}{dp_Td\phi}$ & 12
       & CMS 13 TeV, & 137.1 fb$^{-1}$~\cite{CMS:2021rym}
       \\[0mm]
&  $WW\rightarrow e^\pm \mu^\mp+\sla{E}_T\; (0j)$
&  $m_T$ & 17 (15) & 
ATLAS 13 TeV, &36.1 fb$^{-1}$~\cite{Aaboud:2017gsl} \\[0mm]
& $WZ\rightarrow \ell^+\ell^{-}\ell^{(\prime)\pm}$ 
&  $m_{T}^{WZ}$ & 6 
& ATLAS 13 TeV, &36.1 fb$^{-1}$~\cite{ATLAS:2018ogj} \\[0mm]
& $ Z jj \to \ell^+\ell^- jj$ & $\frac{d\sigma}{d\phi}$ & 12 & ATLAS 13 TeV,
& 139 fb$^{-1}$~\cite{ATLAS:2020nzk} \\[0mm]
& $WW\rightarrow \ell^+\ell^{(\prime) -}+\sla{E}_T\; (1j)$ &
$\frac{d\sigma}{dm_{\ell^+\ell^-}}$ & 10 & ATLAS 13 TeV,
 & 139 fb$^{-1}$~\cite{ATLAS:2021jgw} \\[0mm]
\hline
\hline
  \hline
&$H\to \tau^+\tau^-,W^+W^-, b\bar b  (VBF,ttH+tH)$
&  SS &  7& ATLAS at 13 TeV [Figs. 5,6] & 36.1--139~\cite{ATLAS:2020qdt} \\
&$H\to\gamma Z$ &  SS & 1 & ATLAS at 13 TeV &139 ~\cite{ATLAS:2020qcv}\\
  \multirow{7}{*}
           {\begin{rotate}{90} HIGGS\end{rotate}        }
           &$H\to\mu^+\mu^-$& SS & 1 & ATLAS at 13 TeV~& 139~\cite{ATLAS:2020fzp} \\
&$H\to \gamma\gamma, ZZ,b\bar{b}  (VH)
$ &STXS &43 &  ATLAS at 13 TeV & 139~\cite{ATLAS:2020naq}\\
&$H\to ZZ,b\bar b,\tau^+\tau^- (VH,ttH),W^+W^-(ggH,VBF,ttH)$
& SS & 16  &   CMS at 13 TeV [Table 5]&  35.9--137~\cite{CMS:2020gsy} \\[0mm]
&$H\to\gamma Z
$ &  SS & 1 & CMS at 13 TeV &139~\cite{CMS:2021wog}\\
&$H\to\gamma\gamma$ & STXS & 17 &  
CMS at 13 TeV & 137~\cite{CMS:2021kom}\\[0mm]
&$H\to\tau^+\tau^- \; (ggH, VBF)$  & STXS &11 & 
  CMS at 13 TeV & 137~\cite{CMS:2020dvp}\\[0mm]
&  $H\to W^+W^-\; (VH)$ & STXS & 4 & CMS at 13 TeV & 137~\cite{CMS:2021ixs}\\[0mm]
    \hline
\end{tabular}
\caption{Diboson and Higgs data from LHC used in the analyses.  For
  the $W^+W^-$ results from ATLAS run 2~\cite{Aaboud:2017gsl} we
  combined the data from the last three bins into one to ensure
  gaussianity.}
\label{tab:tgv-h-data}
\end{table}


We also study the implications for HEFT of the available Higgs
data. We consider the Higgs kinematic distributions for the channels
that are available in the Simplified Template Cross Section (STXS)
format, otherwise we use the total signal strength (SS) results. We
summarize in the lower part of Table~\ref{tab:tgv-h-data} the Higgs
data we take into account, specifying its STXS or SS format.  Let us
notice that the correlations among the CMS STXS data for the different
final states is not publicly available.  These are expected to be
important for the channels $\gamma\gamma$ and $\ell\ell\ell\ell$. So,
to make the analysis more robust we have conservatively chosen not to
include the CMS STXS data for $\ell\ell\ell\ell$ for which we only
consider the total SS for this final state.\medskip

We evaluate the theoretical predictions for the Higgs production by
gluon fusion in the channels tagged as STXS in
Table~\ref{tab:tgv-h-data} using
\textsc{MadGraph5\_aMC@NLO}~\cite{Hirschi:2015iia} with the SMEFT@NLO
UFO files~\cite{Degrande:2020evl}. Furthermore, the STXS 1.2
classification was performed using
\textsc{Rivet}~\cite{Buckley:2010ar}.  \medskip

The statistical comparison of the HEFT predictions for the Higgs run 2
data is carried out using the $\chi^2_{\rm Higgs}$ function
\begin{equation}
  \chi^2_{\rm Higgs} \left (\Delta a_C \;,\; a_4 \;,\; a_5 \;,\; a_{17} \;,\;
  a_B \;,\; a_W \;,\; a_G \;,\;
  Y^{(1)}_t \;,\;  Y^{(1)}_b \;,\;  Y^{(1)}_\tau \;,\;  Y^{(1)}_\mu
\right) 
\label{eq:Lhiggs}
\end{equation}
which depends on 11 Wilson coefficients. Once again, we do not take
into account the contributions from the anomalous gauge boson
couplings to fermion pairs due to the stringent bounds emanating from
the EWPO data on these couplings.  \medskip

\section{Results}
\label{sec:results}

As discussed above, in HEFT the non-linear realization of the gauge
symmetry allows for independent statistical analyses of the datasets
involving corrections to the gauge-boson-fermion and gauge-boson self
couplings (EWPO and TGC) and the Higgs interactions since the
couplings impacting these two sectors are not connected. \medskip

\subsection{Constraints from EWPO}

\begin{figure}[h!]
  \centering
  \includegraphics[height=0.9\textheight]{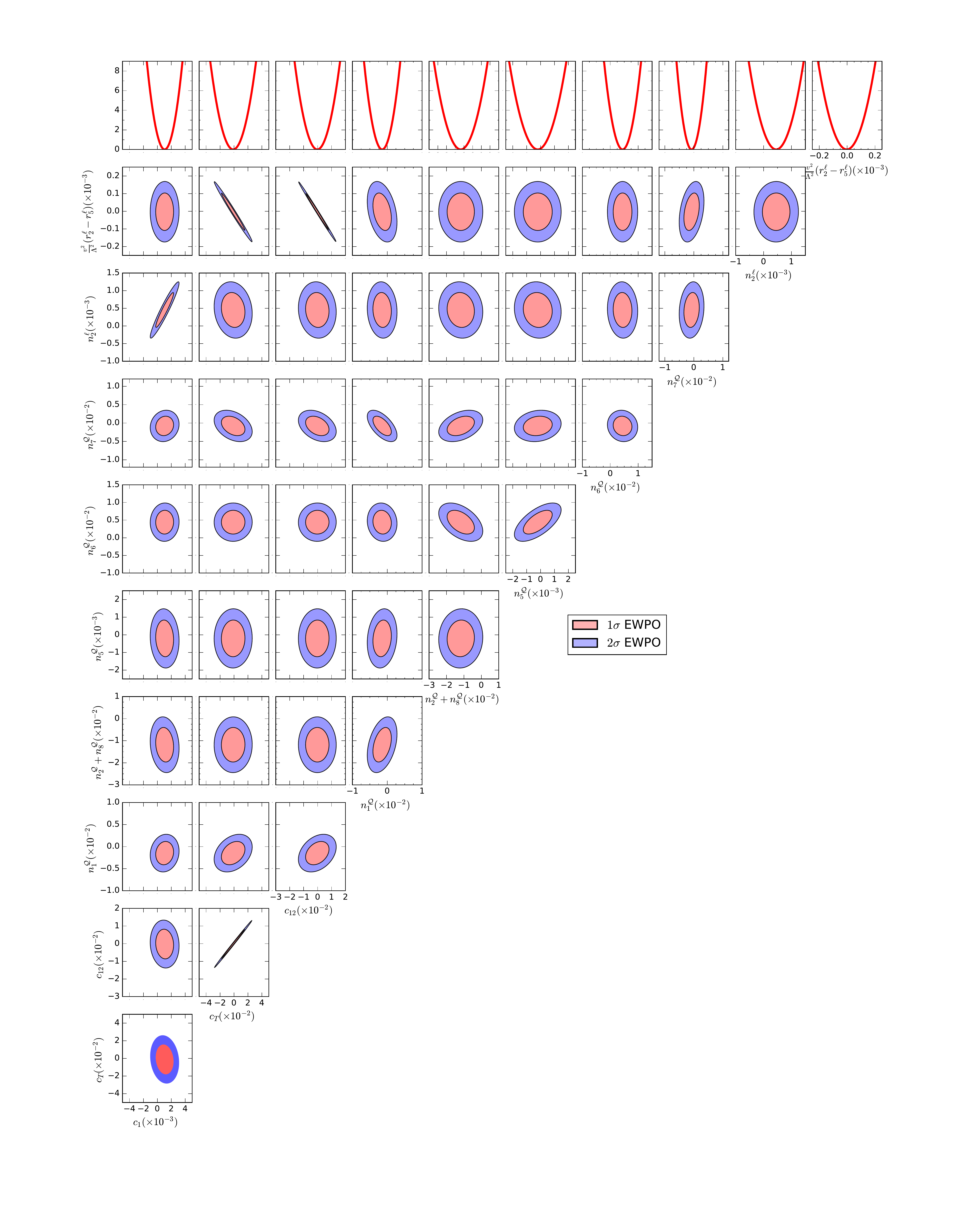}
  \caption{One- and two-dimensional marginalized projections of
    $\Delta\chi^2_{\rm EWPO}$ for the Wilson coefficients $ c_1$,
    $c_T$, $c_{12}$, $n_1^Q$, $n_2^\cQ + n^\cQ_4$, $n_5^Q$, $n_6^Q$,
    $n_7^Q$, $n_2^\ell$, and $r^\ell_2-r^\ell_5 $, as indicated in the
    panels after marginalizing over the remaining fit parameters.}
  \label{fig:ewpd-cont}
\end{figure}

We start our analysis focusing on the HEFT operators in
Eq.~\eqref{eq:oper-ewpd} that contribute to the EWPO.  The results are
graphically presented in Fig.~\ref{fig:ewpd-cont} which depicts the
one- and two-dimensional projections of $\Delta\chi^2_{\rm EWPO}$ as a
function of the Wilson coefficients after marginalizing over those non
displayed in each panel. \medskip

More quantitatively, the corresponding best values, uncertainties,
95\% CL ranges, and correlations are presented in
Eq.~\eqref{eq:ewpdcorr}.  As the EWPO analysis was performed including
only the linear contribution on the Wilson coefficients (here denoted
generically $f_i$) to the observables, by definition $\Delta\chi^2$
takes the form
\begin{equation}  
\Delta\chi^2=\sum_{i=1}^{N}
  \left({f_i}-{f^0_i}\right)
  \,V^{-1}_{ij}  \left({f_j}-{f^0_j}\right)\;,
\label{eq:chilin}
\end{equation}
where $f^0_j$ defines  the best fit point and $V$ is the covariance matrix
\[
  V_{ij}\equiv\sigma_i\,\sigma_j\,\rho_{ij}\;,
\]
with $\sigma_j$ being the uncertainties and $\rho_{ij}$ the
correlation matrix. For the EWPO analysis we find the best fit values,
uncertainties and correlation matrix:
\begin{equation}
\tiny{  \begin{array}{|c@{\hskip 0.3cm}|c|cccccccccc|}
     \hline 
     \multirow{3}{*}{ }
&     & c_1 & c_T& c_{12}& n_1^\cQ & n_2^\cQ+n_8^\cQ & n_5^\cQ & n_6^\cQ & 
     n_7^\cQ & n_2^\ell & (r_2^\ell-r_5^\ell)\frac{v^2}{\Lambda^2}
     \\[+0.2cm]\cline{2-12}
&  {\rm b.f.}\; (\times 10^{-3})
  &  1.1 & -1.2 & -0.21 & -1.5 & -12 & -0.20 & 4.4  & -0.78 & 0.45 & -0.0027
     \\[+0.2cm]\cline{2-12}
&\sigma \; (\times 10^{-3})
&0.85 & 11. & 5.4 & 1.7 & 5.2 & 0.67 & 2.2  & 1.7 & 0.33 & 0.069
\\[+0.2cm]\cline{2-12}
&  95\%\;{\rm CL}\; (\times 10^{-3})
&(-0.66,2.7)& (-23,21) &(-11,11) & (-4.9,2.0)&(-22,-1.5) &(-1.2,1.2)&
(-0.024,8.8)& (-4.2,2.7)&(-0.2,1.1)&(-0.14,0.13)
     \\[+0.2cm]\hline     
     \multirow{10}{*}{$\rho$}
&c_1 & 1.000 & -0.136 & -0.074 & 0.108 & -0.128 & -0.075 & 0.019 & 0.113 & 0.950 & 0.006 \\
&c_T&     
     -0.136 & 1.000 & 0.998 & 0.325 & 0.009 & 0.012 & 0.004 & -0.362 & -0.130 & -0.991 \\
&c_{12}&     
     -0.074 & 0.998 & 1.000 & 0.333 & -0.002 & 0.009 & 0.008 & -0.359 & -0.071 & -0.997 \\
&n_1^\cQ &      
     0.108 & 0.325 & 0.333 & 1.000 & 0.385 & 0.146 & -0.077 & -0.602 & 0.138 & -0.346 \\
&n_2^\cQ+n_8^\cQ &      
     -0.128 & 0.009 & -0.002 & 0.385 & 1.000 & 0.042 & -0.467 & 0.384 & -0.047 & -0.001 \\
&n_5^\cQ &      
     -0.075 & 0.012 & 0.009 & 0.146 & 0.042 & 1.000 & 0.640 & 0.146 & -0.098 & -0.001 \\
&n_6^\cQ &       
     0.019 & 0.004 & 0.008 & -0.077 & -0.467 & 0.640 & 1.000 & -0.077 & -0.045 & 0.000 \\
&n_7^\cQ &           
     0.113 & -0.362 & -0.359 & -0.602 & 0.384 & 0.146 & -0.077 & 1.000 & 0.142 & 0.347\\
&n_2^\ell &     
     0.950& -0.130 & -0.071 & 0.138 & -0.047 & -0.098 & -0.045 & 0.142 & 1.000 & 0.006 \\
&(r_2^\ell-r_5^\ell)\frac{v^2}{\Lambda^2} &
0.006 & -0.991 & -0.997 & -0.346 & -0.001 & -0.001 & 0.000 & 0.347 & 0.006 & 1.000\\\hline\end{array}}.
\label{eq:ewpdcorr}
\end{equation}

Altogether, the overall picture is that there is a good agreement with
the standard model.  As discussed in Ref.~\cite{Brivio:2016fzo}, there
are two main differences with respect to the corresponding analysis to
EWPO obtained assuming SMEFT with operators up to dimension six, which
are straight forward to identify by working in the Hagiwara, Ishihara,
Szalapski, and Zeppenfeld (HISZ) basis~\cite{Hagiwara:1993ck,
  Hagiwara:1996kf} (see for example Refs.~\cite{Pomarol:2013zra,
  Corbett:2017qgl, Alves:2018nof, Almeida:2021asy}): (i) in the SMEFT
no contribution to the $U$ parameter is generated at dimension six
while in the HEFT $c_{12}$ gives contribution to $U$; (ii) assuming
universality in the gauge-fermion couplings, the $W$ and $Z$ couplings
to fermions are linked in the SMEFT and receive contributions from the
coefficients of five non-oblique operators while in the HEFT an
additional non-oblique operator coefficient ($n_7^\cQ$) allows for
independent variations of the $W$ and $Z$ couplings to
quarks. \medskip

As seen in Fig.~\ref{fig:ewpd-cont} and Eq.~\eqref{eq:ewpdcorr}, a
nonzero $c_{12}$ has most effect on the allowed ranges of $c_T$ and
$(r_2^\ell-r_5^\ell)\frac{v^2}{\Lambda^2}$, which are the Wilson
coefficients with which $c_{12}$ is mostly correlated.  This is due to
possible cancellations between the oblique contributions and
$\delta G_F$ in the $Z$ observables and $\Delta M_W$; see
Ref.~\cite{Brivio:2016fzo} for further details.  Notice that these
cancellations are possible at NLO in the HEFT but not in the SMEFT at
dimension six~\cite{Brivio:2016fzo}.  This results into a weakening of
the bounds on $c_T$ and $(r_2^\ell-r_5^\ell)\frac{v^2}{\Lambda^2}$ by
about a factor $\sim 20$, though they still remain constrained at the
percent and per mil level respectively. \medskip

Conversely, the contribution of the quark-current operator $\nq_7$ has
a more modest quantitative impact. In fact, it is still the case that
the EWPO analysis favors non-vanishing values for the coefficients
contributing to the down-quark couplings to the $Z$, in particular
$(\nq_2+\nq_8)$, at $2\sigma$. This is a well-known result driven by
the $2.7\sigma$ discrepancy between the observed $A_{FB}^{0,b}$ and
the SM.  \medskip

\subsection{Triple gauge couplings constraints}

\begin{figure}[h!]
  \centering
  \includegraphics[width=0.75\textwidth]{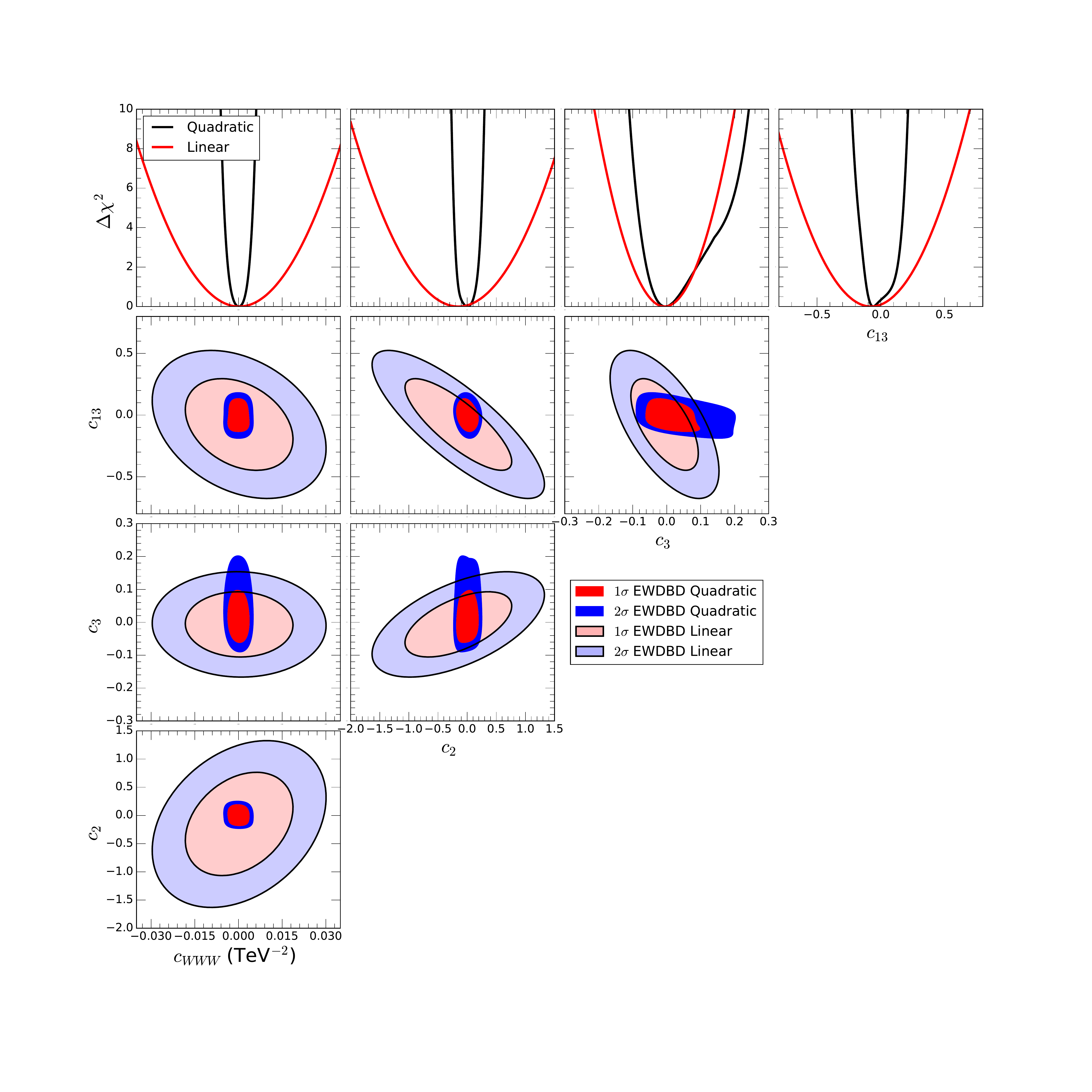}
  \caption{One- and two-dimensional marginalized projections of
    $\Delta\chi^2_{\rm TGC}$ for the Wilson coefficients $c_2$, $c_3$,
    $c_{WWW}$, and $c_{13}$ as indicated in the panels after
    marginalizing over the remaining fit parameters. The results are
    shown for analyses including only the linear contributions of the
    Wilson coefficients (red curves and lighter regions) as well as up
    to quadratic contributions (black lines and darker regions).}
  \label{fig:tgc-cont}
\end{figure}

The results of our analyses of the EWDBD are graphically presented in
Fig.~\ref{fig:tgc-cont}.  As mentioned in Sec.~\ref{sec:frame}, for
the sake of simplicity in our EWDBD analysis, we discard possible
effects from anomalous couplings of gauge bosons to fermion pairs,
which are well constrained by the EWPO data, and focus on the
constraints on the bosonic operators $\cP_2$, $\cP_3$, $\cP_{13}$, and
$\cP_{WWW}$. \medskip

We perform the EWDBD analyses considering only the linear
contributions of the Wilson coefficients to the observables as well as
including up to quadratic contributions. At linear order
$\Delta \chi^2_{\rm TGC}$ takes the form of Eq.~\eqref{eq:chilin} with
best fit, uncertainties and correlations
\begin{equation}
\begin{array}{|c@{\hskip 0.3cm}|c|cccc|}\hline
&   & c_2 & c_3 & c_{13} & c_{WWW}/({\rm TeV}^{-2}) 
\\[+0.2cm]\cline{2-6}     
& {\rm b.f.}\; & -0.15 & -0.0063 & -0.076 & 0.00024
\\[+0.2cm]\cline{2-6}     
& \sigma     & 0.60 & 0.065 & 0.24 & 0.012 
\\\hline
\multirow{4}{*}{$\rho$}
&  c_2 & 1.000 & 0.515 & -0.855 & 0.275 \\
&c_3& 0.515 & 1.000 & -0.675 & -0.075 \\
& c_{13}& -0.855 & -0.675 & 1.000 & -0.365 \\
& \frac{c_{WWW}}{({\rm TeV}^{-2})}& 
0.275 & -0.075 & -0.365 & 1.000 \\\hline
\end{array}\;.
\label{eq:tgvcorr}
\end{equation}
The corresponding 95\%CL allowed ranges for both analysis are listed
in Table~\ref{tab:ranges}. \medskip

From Fig.~\ref{fig:tgc-cont} or Eq.~\eqref{eq:tgvcorr} we can see that
$c_2$, $c_3$ and $c_{13}$ are strongly correlated in the linear
analysis due to their contributions to the $ZWW$ and $\gamma WW$
vertices through $\Delta\kappa_Z$ and $\Delta\kappa_\gamma$; see
Eq.~\eqref{eq:kappa}. It is interesting to notice that the $W\gamma$
production plays a significant role in constraining $c_{WWW}$ at
linear order due to the use of kinematic distributions designed to
avoid the cancellation of its linear
contribution~\cite{Azatov:2017kzw, Azatov:2019xxn}.\medskip

When compared to the corresponding EWDBD analysis performed in the
framework of SMEFT at dimension six in the HISZ
basis~\cite{Hagiwara:1993ck, Hagiwara:1996kf}, the main difference is
the contribution of the operator $\cP_{13}$ which has no linear
sibling at dimension six. As seen in Fig.~\ref{fig:ewpd-cont}, the
presence of this additional coefficient leads to the relaxation of the
bounds on $c_2$ (and less significantly on $c_3$) obtained when
considering only linear contributions, as a consequence of the
mentioned correlations.  We also see that once the effects of the
operators are included to quadratic order in the coefficients, the
inclusion of $c_{13}$ in the analysis has minimal impact on the
determination of the other three coefficients. Furthermore, the
quadratic analysis leads to stronger limits by a factor $3$--$5$; see
Table~\ref{tab:ranges}.\medskip

\begin{table} [ht]
  \begin{tabular}{|c||c|c|}
  \hline 
&\multicolumn{2}{c|}{95\% CL Range}\\\cline{2-3}
  & Linear & Quadratic \\\hline
  $c_2$ & $(-1.4,1.0)$&$(-0.21,0.23)$\\
  $c_3$ & $(-0.14,0.12)$&$(-0.080,0.16)$\\
  $c_{13}$ & $(-0.57,0.41)$&$(-0.16,0.16)$\\
  $c_{WWW}/{\rm TeV}^{-2}$ & $(-2.4,2.5)\times 10^{-2}$& $(-4.6,4.6)\times 10^{-3}$\\
  $a_4$ & $(-0.17,1.10)$& $(-0.12,0.44)$\\
  $a_5$ & $(-0.38,0.83)$& $(-0.70,0.71)$\\
  $a_{17}$& $(-0.38,0.35)$& $(-0.49,0.30)$\\
  $a_B$ &$(-0.71,3.2)\times 10^{-2}$&$(-0.53,2.4)\times 10^{-2}$\\
  $a_W$ &$(-10,2.3)\times 10^{-2}$&$(-5.5,1.8)\times 10^{-2}$\\
  $a_G$ &$(-1.1,0.37)\times 10^{-3}$&$(-1.2,0.38)\times 10^{-3}$\\
 $\Delta a_C$  &$(-0.20,0.046)$& $(-0.17,0.062)$\\
  $Y_t^{(1)}/Y_t^{(0)}-1$
  &$(-0.17,0.29)$& $(-0.14,0.28)$\\
  ${Y_b^{(1)}}/{Y_b^{(0)}}-1$ &
  $(-0.50,0.15)$&    $(-2.1,-1.6)\,\cup\, (-0.42,0.13)$\\
  ${Y_\tau^{(1)}}/{Y_\tau^{(0)}}-1$ & 
  $(-0.37,0.062)$&    $(-2.0,-1.6)\,\cup\, (-0.35,0.026)$\\
    ${Y_\mu^{(1)}}/{Y_\mu^{(0)}}-1$ & 
 $(-0.60,0.54)$&    $(-2.4,-1.1)\,\cup\, (-0.87,0.39)$\\\hline
\end{tabular}
\caption{Marginalized 95\% CL allowed ranges for the Wilson coefficients
of the operators constrained by the analysis of LHC EWDBD and Higgs results.}
\label{tab:ranges}
\end{table}

\subsection{Higgs couplings}

\begin{figure}[h!]
  \centering
  \includegraphics[height=0.95\textwidth]{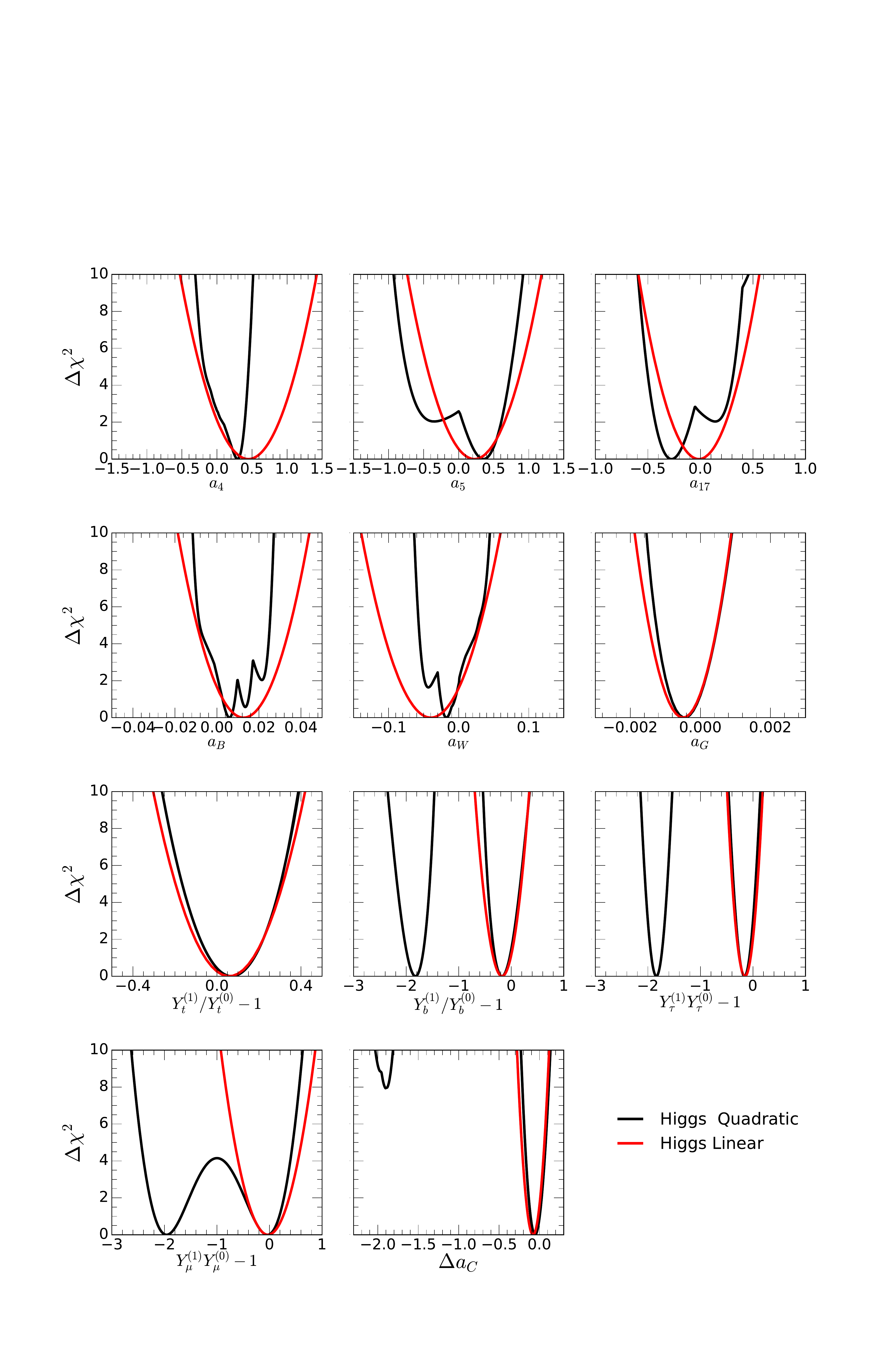}
  \caption{$\Delta \chi^2$ as a function of the Wilson coefficients
    $a_4 \;,\; a_5 \;,\; a_{17} \;,\; a_B \;,\; a_W \;,\; a_G \;,\;
    Y^{(1)}_t \;,\; Y^{(1)}_b \;,\; Y^{(1)}_\tau \;,\; Y^{(1)}_\mu\;,$
    and $\Delta a_C$ as indicated in the panels after marginalizing
    over the remaining fit parameters. The red (black) line stands for
    the analysis considering the linear (and quadratic) contributions
    of the Wilson coefficients.}
  \label{fig:higgs-1d}
\end{figure}

\begin{figure}[h!]
\centering
\includegraphics[width=0.4\textwidth]{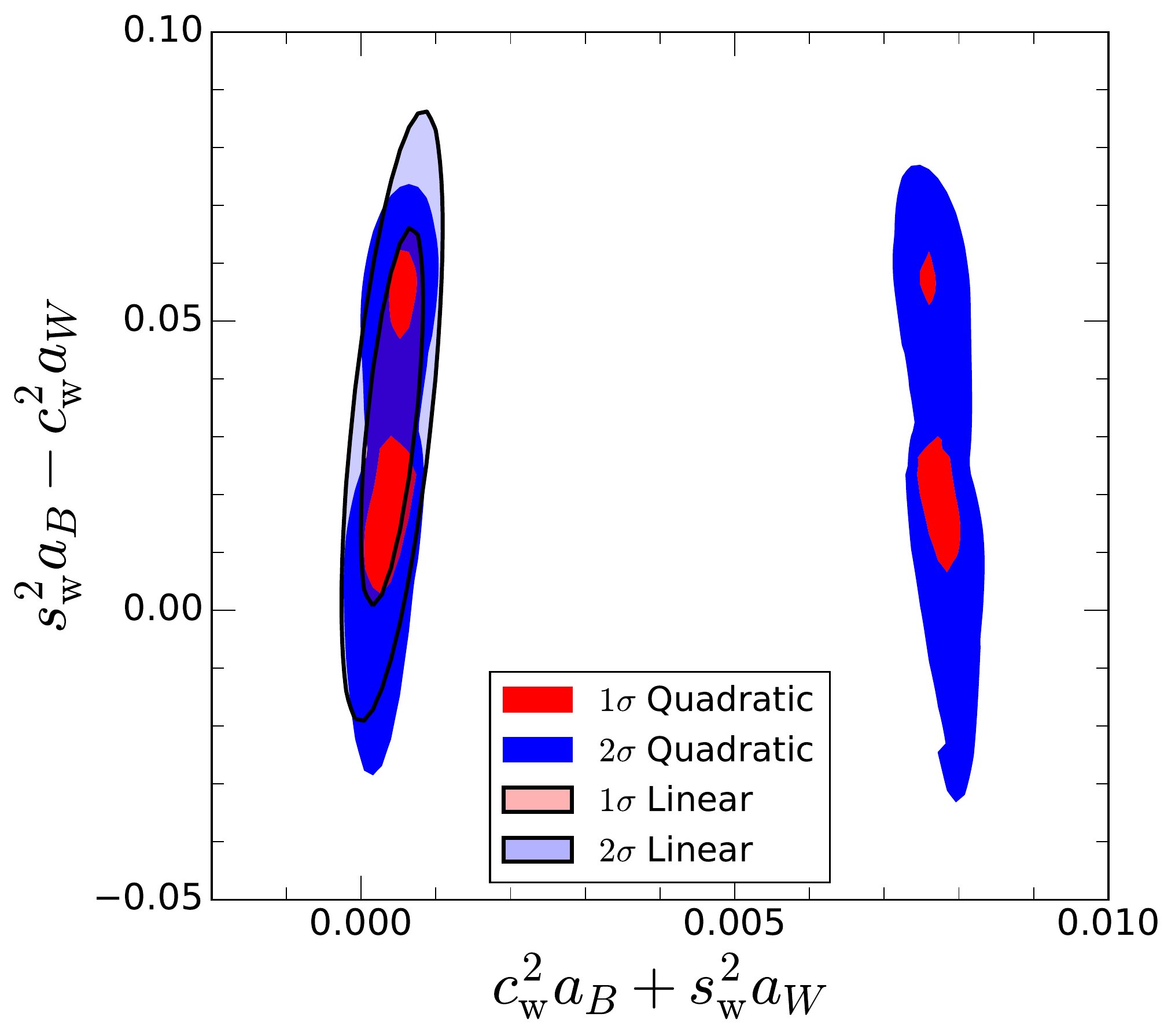}
\caption{$1\sigma$ and 95\% CL (2dof) allowed regions from the Higgs
  analysis for the combinations $c_{\rm W}^2 a_B+ s_{\rm W}^2 a_W$ and
  $s_{\rm W}^2 a_B- c_{\rm W}^2 a_W$.  The results are shown for the
  analyses including only the linear contributions of the Wilson
  coefficients (lighter regions) as well as up to quadratic
  contributions (darker regions).}
  \label{fig:contgaga}
\end{figure}

\begin{figure}[h!]
\centering
\includegraphics[width=\textwidth]{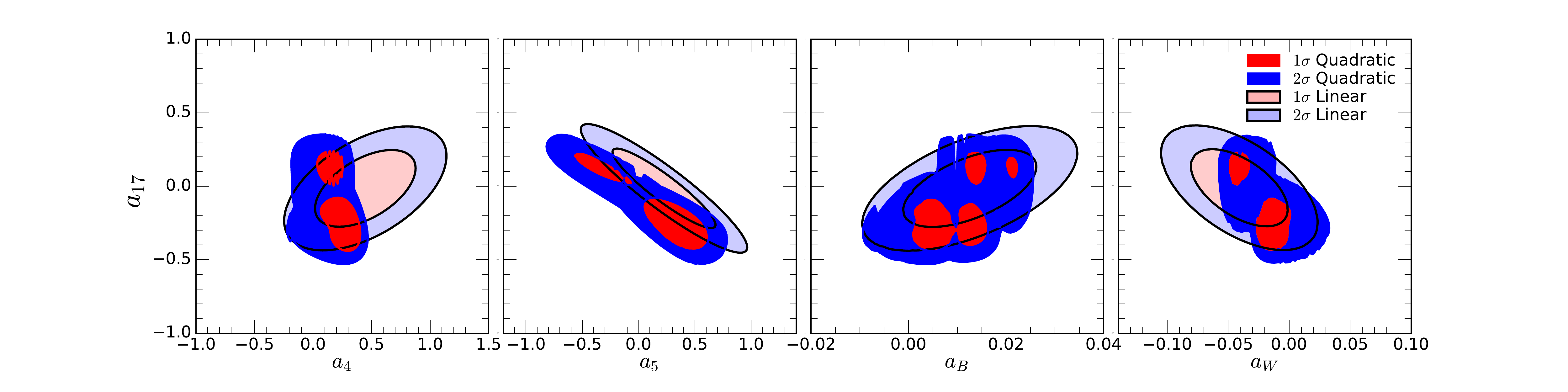}
\caption{$1\sigma$ and 95\% CL (2dof) allowed regions from the Higgs
  analysis for $a_{17} \otimes $ ( $a_4$, $a_5$, $a_B$, and $a_W$),
  profiling over the undisplayed parameters.  The results are shown
  for analyses including only the linear contributions of the Wilson
  coefficients (lighter regions) as well as up to quadratic
  contributions (darker regions).}
  \label{fig:conta17}
\end{figure}

The results of our analysis of the Higgs results are graphically
presented in Figs.~\ref{fig:higgs-1d},~\ref{fig:contgaga},
and~\ref{fig:conta17}.  As mentioned in Sec.~\ref{sec:frame}, we
assume that $\bar a_{T(1)}$ couplings are at most ${\cal O}(1)$ and
neglect the contributions from $a_{T(1)}=c_{T(1)} \,\bar a_{T(1)}$ to
the Higgs observables after imposing the strong constraints on
$c_{T (1)}$ from EWPO. Thus, we perform an analysis in terms of the 11
coefficients in Eq.~\eqref{eq:Lhiggs}.\medskip
%

We have carried out the Higgs analyses considering only the linear
contributions of the Wilson coefficients to the observables as well as
including up to quadratic contributions. The one-dimensional
projections are depicted in Fig.~\ref{fig:higgs-1d}. First of all,
notice that the results are compatible with the SM at 95\% CL in both
the linear and the quadratic analyses. Moreover, the figure shows that
the dominant source of degeneracy that remains in these parameters
using the LHC run 2 data are those related to the Yukawa couplings
$Y^{(1)}_f$ for $f=b$, $\tau$ and $\mu$, which are associated to the
reversing of the SM Yukawa coupling sign; see Eq.~\eqref{eq:yukdeg}.
As seen in this figure, for the quadratic analysis $\Delta a_C$
presents a second minima around $\Delta a_C\sim 2$ which is associated
with the change of sign of the $H V^\mu V_\mu$ in Eq.~\eqref{eq:vvdeg}
but it lays at $\Delta\chi^2\sim 8$.  This degeneracy is mainly broken
by the $tH$ data which receives contribution from both $HVV$ and
$Ht\bar t$ vertices while only the first one changes sign for
$\Delta a_C\sim 2$. Correspondingly the degeneracy for
$Y^{(1)}_t=-Y^{(0)}_t$ is also broken and in this case the solution
with inverted SM sign lays at $\Delta\chi^2\gg 10$. This is so because
the top yukawa coupling also contributes to the gluon-gluon-Higgs
production in Eq.~\eqref{eq:glugludeg} and it can be resolved by the
STXS data.  Similarly, no degenerate solution is found for $a_G$.  In
summary, the larger available luminosity as well as Higgs kinematic
distributions eliminates the degeneracies associated to $a_G$ and
$Y^{(1)}_t$ that were observed
previously~\cite{Brivio:2013pma}. \medskip

Finally, there remain the quasi-degenerate solutions associated to the
$\cP_W$ and $\cP_B$ corrections to the $H\,\gamma\,\gamma$ vertex in
Eq.~\eqref{eq:gagadeg} for which the data only constrain its
modulus. This is displayed in Fig.~\ref{fig:contgaga} where we show
the marginalized two-dimensional projections of
$\Delta\chi^2_{\rm Higgs}$ for the coefficients $a_W$ and $a_B$. For
the sake of clarity, we plot the allowed regions for the combinations
$c_{\rm W}^2 a_B+ s_{\rm W}^2 a_W$ (which corrects
$HF^{\mu\nu} F_{\mu\nu}$ ) and $s_{\rm W}^2 a_B- c_{\rm W}^2 a_W$
(which gives a contribution to $HZ^{\mu\nu}Z_{\mu\nu}$). In the figure
we clearly see the two SM-line solutions around
$c_{\rm W}^2 a_B+ s_{\rm W}^2 a_W\sim 0$, and
$c_{\rm W}^2 a_B+ s_{\rm W}^2 a_W\sim v\,G^{\gamma\gamma}_{SM}\simeq
0.0084$. \medskip

%

One thing to notice is that, when the analysis is preformed at linear
order, $\Delta \chi^2_{\rm Higgs}$ takes the form of
Eq.~\eqref{eq:chilin} with best fit, uncertainties and correlations\\[+0.3cm]
\begin{equation}
\footnotesize{
  \begin{array}{|c@{\hskip 0.3cm}|c|ccccccccccc|}       
     \hline 
& & a_4 &a_5& a_{17}& a_B & a_W & a_G& \Delta a_C & 
     \frac{Y_t^{(1)}}{Y_t^{(0)}}-1& \frac{Y_b^{(1)}}{Y_b^{(0)}}-1
     &\frac{Y_\tau^{(1)}}{Y_\tau^{(0)}}-1 & \frac{Y_\mu^{(1)}}{Y_\mu^{(0)}}-1
\\[+0.2cm]\cline{2-13}
&  {\rm b.f.}\; &
0.45 &   0.22 &  -0.015 &   0.013 &  -0.040 &  -0.00050 &  -0.080 &   0.058 &  -0.17 &  -0.15 &  -0.028 \\[+0.2cm]\cline{2-13}
& \sigma     &    
  0.31 &   0.30 &   0.18 &   0.0099 &   0.031 &   0.00044 &   0.063 &   0.11 &   0.16 &   0.11 &   0.28
    \\\hline
  \multirow{11}{*} {$\rho$}
  & a_4&
1.000 & -0.395 & 0.485 & 0.895 & -0.985 & 0.025 & -0.605 & -0.155 & -0.575 & -0.495 & -0.225 \\
&a_5 &  
-0.395 & 1.000 & -0.935 & -0.075 & 0.265 & -0.075 & -0.025 & -0.185 & -0.085 & 0.045 & 0.005 \\
  &a_{17}&                                     
0.485 & -0.935 & 1.000 & 0.515 & -0.575 & -0.125 & -0.345 & 0.145 & -0.255 & -0.405 & -0.185 \\
    &a_B&                                       
0.895 & -0.075 & 0.515 & 1.000 & -0.995 & -0.115 & -0.705 & -0.005 & -0.625 & -0.505 & -0.215 \\
&a_W&                                        
-0.985 & 0.265 & -0.575 & -0.995 & 1.000 & -0.075 & 0.575 & 0.105 & 0.425 & 0.435 & 0.145 \\
  &a_G&  
0.025 & -0.075 & -0.125 & -0.115 & -0.075 & 1.000 & 0.235 & -0.735 & 0.005 & 0.145 & -0.085 \\
&\Delta a_C &                              
-0.605 & -0.025 & -0.345 & -0.705 & 0.575 & 0.235 & 1.000 & 0.345 & 0.785 & 0.535 & 0.185 \\
&     \frac{Y_t^{(1)}}{Y_t^{(0)}}-1&  
-0.155 & -0.185 & 0.145 & -0.005 & 0.105 & -0.735 & 0.345 & 1.000 & 0.465 & 0.165 & 0.035 \\
& \frac{Y_b^{(1)}}{Y_b^{(0)}}-1   &  
-0.575 & -0.085 & -0.255 & -0.625 & 0.425 & 0.005 & 0.785 & 0.465 & 1.000 & 0.535 & 0.145 \\
  &\frac{Y_\tau^{(1)}}{Y_\tau^{(0)}}-1 &  
-0.495 & 0.045 & -0.405 & -0.505 & 0.435 & 0.145 & 0.535 & 0.165 & 0.535 & 1.000 & 0.105 \\
&\frac{Y_\mu^{(1)}}{Y_\mu^{(0)}}-1&        
-0.225 & 0.005 & -0.185 & -0.215 & 0.145 & -0.085 & 0.185 & 0.035 & 0.145 & 0.105 & 1.000 \\
\hline
\end{array}}
\label{eq:higgslin}
\end{equation}
\medskip

When compared to the corresponding analysis performed in the framework
of SMEFT at dimension-six in the HISZ basis~\cite{Hagiwara:1993ck,
  Hagiwara:1996kf} the main difference is the contribution of the
operator $\cP_{17}$ which has no linear sibling at dimension six. This
operator leads to vertices $\partial^\mu H\, Z_{\mu\nu} Z^\nu$ and
$\partial^\mu H\, F_{\mu\nu} Z^\nu$ and, generically its addition
enlarges the allowed range for the rest of the coefficients
contributing to the Higgs-gauge-boson trilinear interactions.
Therefore, it is mostly correlated with those coefficients which
modify the $HZZ$ and $H\gamma Z$ couplings as well, {\em i.e.}  $a_4$,
$a_5$, $a_B$, and $a_W$, as can be seen in the corresponding entries
in Eq.~\eqref{eq:higgslin} and it is graphically illustrated in
Fig.~\ref{fig:conta17}. \medskip

\section{Discussion}
\label{sec:disc}

In this work we have presented the results of comprehensive analyses
of low-energy electroweak precision measurements as well as LHC data
on gauge boson pair production and Higgs observables in the context of
the effective low energy theory for a dynamical Higgs. We focused on
observables related to the electroweak sector, which at present allow
for precision tests of the couplings between electroweak gauge bosons
and fermions, triple electroweak gauge couplings and the couplings of
the Higgs to fermions and gauge bosons.  For the sake of assessing the
impact of the Higgs kinematic distributions, we performed an analysis
including the most updated STXS Higgs data in combination with the
Higgs total signal strengths for those channels for which no kinematic
information is available.  In total, the analyses of EWPO and EWDBD
and Higgs results from LHC run 2 encompass 15+73+101=189 observables;
see Sec.~\ref{sec:frame} for further details. \medskip

We worked in the framework of effective Lagrangians assuming the
non-linear (chiral) realization of the electroweak gauge symmetry, the
so-called HEFT which we have considered up to next-to-leading order.
Under the flavor assumption that the new operators do not introduce
additional tree level sources of flavor violation nor violation of
universality of the weak current, the analysis involves a total of 25
Wilson coefficients, of which 10 are more severely constrained by
EWPO, 4 in addition are constrained by EWDBD, and 11 are determined by
the Higgs data analysis. \medskip

All of the analyses performed show no statistically significant source
of tension with the SM. We find
\begin{eqnarray}
\chi^2_{\rm min\;EWPO,\; SM}= 18.3 \;,
&&{\rm\; 15\;observables\,,} \nonumber
\\
\chi^2_{\rm min\;EWDBD\; SM}= 63.7\;,  &&
    {\rm\; 73\;observables\,,}
  \\
  \chi^2_{\rm min\;Higgs,\; SM}= 99.2\;,
  && 
{\rm 101\;observables\,},   \nonumber
\end{eqnarray}
to be compared with 
\begin{eqnarray}
\chi^2_{\rm min\;EWPO,\; HEFT\; Linear}=6 \; ,&&
{\rm\; 15\;observables\;\&\; 10\; coefficients\,,}\nonumber
\\
\chi^2_{\rm min\;EWDBD, HEFT\,Linear\; [HEFT\,Quadratic]}= 63\;[63 ], &&
{\rm\; 73\;observables\;\& \; 4\; coefficients\,,} 
  \\
\chi^2_{\rm min\;Higgs, HEFT\,Linear\; [HEFT\,Quadratic]}= 89\;[87 ], &&
{\rm\; 101\;observables\;\& \; 11\; coefficients\,.} 
   \nonumber
\end{eqnarray}

\begin{figure}[ht!]
\centering
\includegraphics[width=0.49\textwidth]{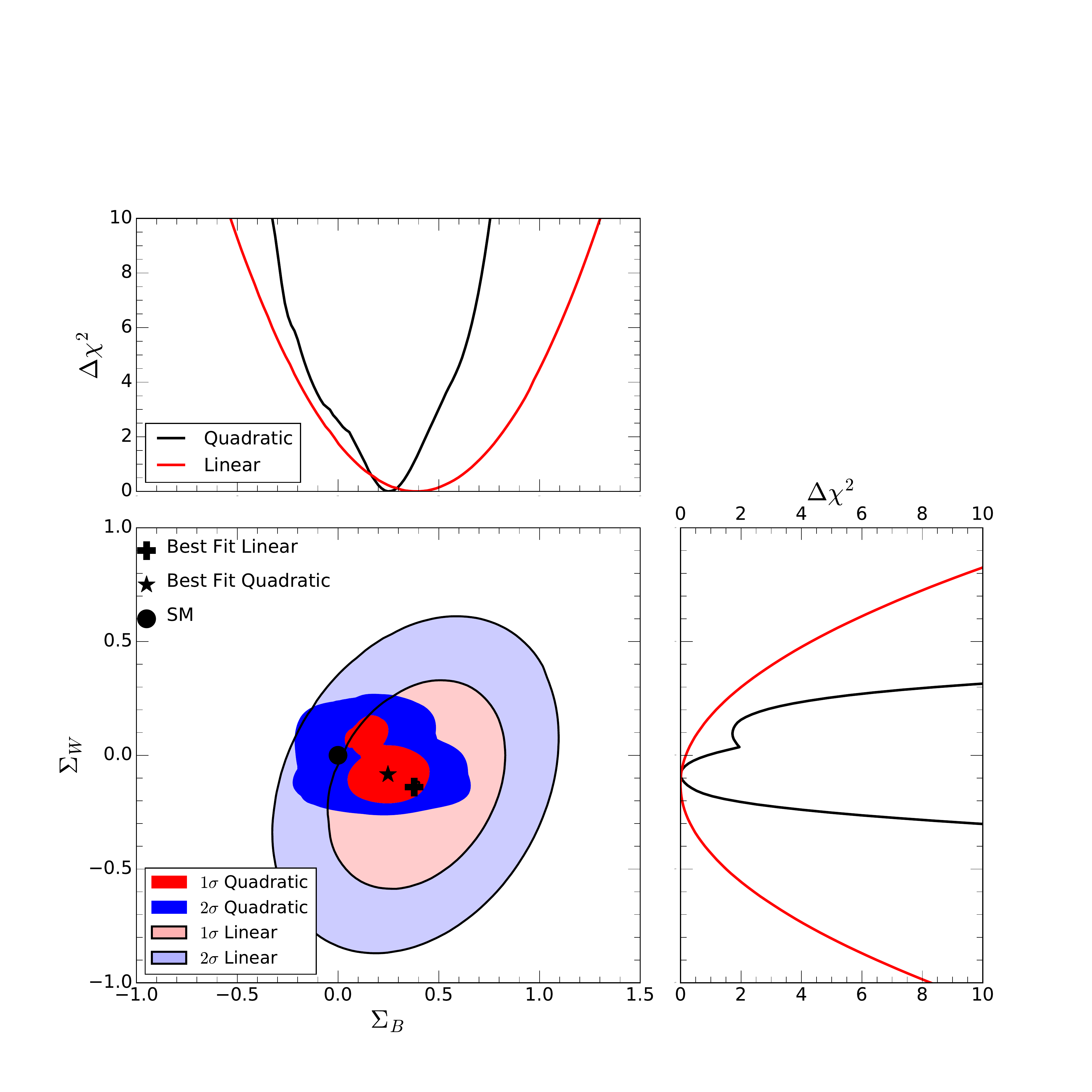}
\includegraphics[width=0.49\textwidth]{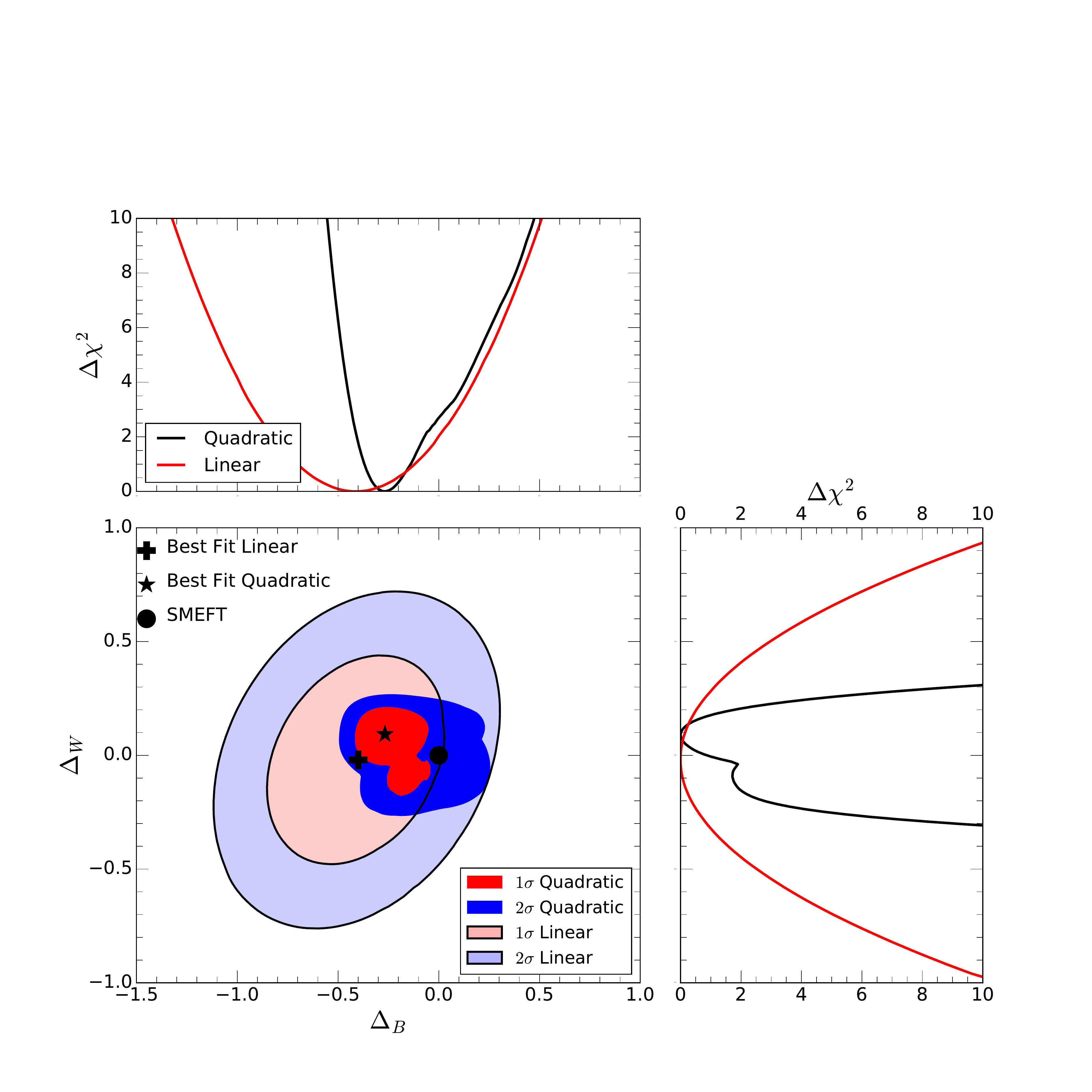}
\caption{Present bounds on $\Sigma_B$, $\Sigma_W$, $\Delta_B$ and
  $\Delta_W$ (see text for the details on their definition) as
  obtained from the most recent combined global analysis of Higgs and
  EWDBD after profiling over the undisplayed parameters spanned in the
  analysis ($\Delta a_C$, $a_B$, $a_G$, $a_W$, $a_{17}$, $Y^{(1)}_t$,
  $Y^{(1)}_b$, $ Y^{(1)}_\tau$,$ Y^{(1)}_\mu$, $c_{13}$, and
  $c_{WWW}$).}
\label{fig:sigdel}
\end{figure}

The results of the analysis can be confronted with those performed
under different assumptions for the nature of the Higgs boson {\i.e.}
whether it is an isosinglet or a member of a $SU(2)_L$ doublet.  In
particular, as summarized in Sec.~\ref{sec:thframe}, a particularly
sensitive probe of the nature of the Higgs boson is associated with
the (de)correlation of the contributions to TGC's and
Higgs-gauge-boson couplings in the SMEFT and HEFT scenarios. This
comparison can be quantified in terms of the four parameters in
Eq.~\eqref{eq:sigdel}~\cite{Brivio:2013pma,Brivio:2016fzo}.
Figure~\ref{fig:sigdel} shows the current status of the bounds on the
two relevant planes of these coefficients. \medskip

The constraints of $\Sigma_B$, $\Sigma_W$, $\Delta_B$, and $\Delta_W$
shown in Fig.~\ref{fig:sigdel} present a significant improvement with
respect to the bounds previously shown in Fig.~3 of
Ref.~\cite{Brivio:2016fzo} in the $\Delta_B$ and $\Sigma_B$ axis. They
also show that the comparison between the two scenarios is robust
irrespective of whether the analysis is performed at linear or
quadratic order in the Wilson coefficients. We learn from this figure
that, presently, the data gathered so far is not enough to distinguish
between the two scenarios for the Higgs nature.  From the right panel
we read that the SMEFT lies within the 1$\sigma$ allowed region of
either the linear or quadratic analysis.\medskip

\begin{figure}[ht!]
\centering
\includegraphics[width=0.4\textwidth]{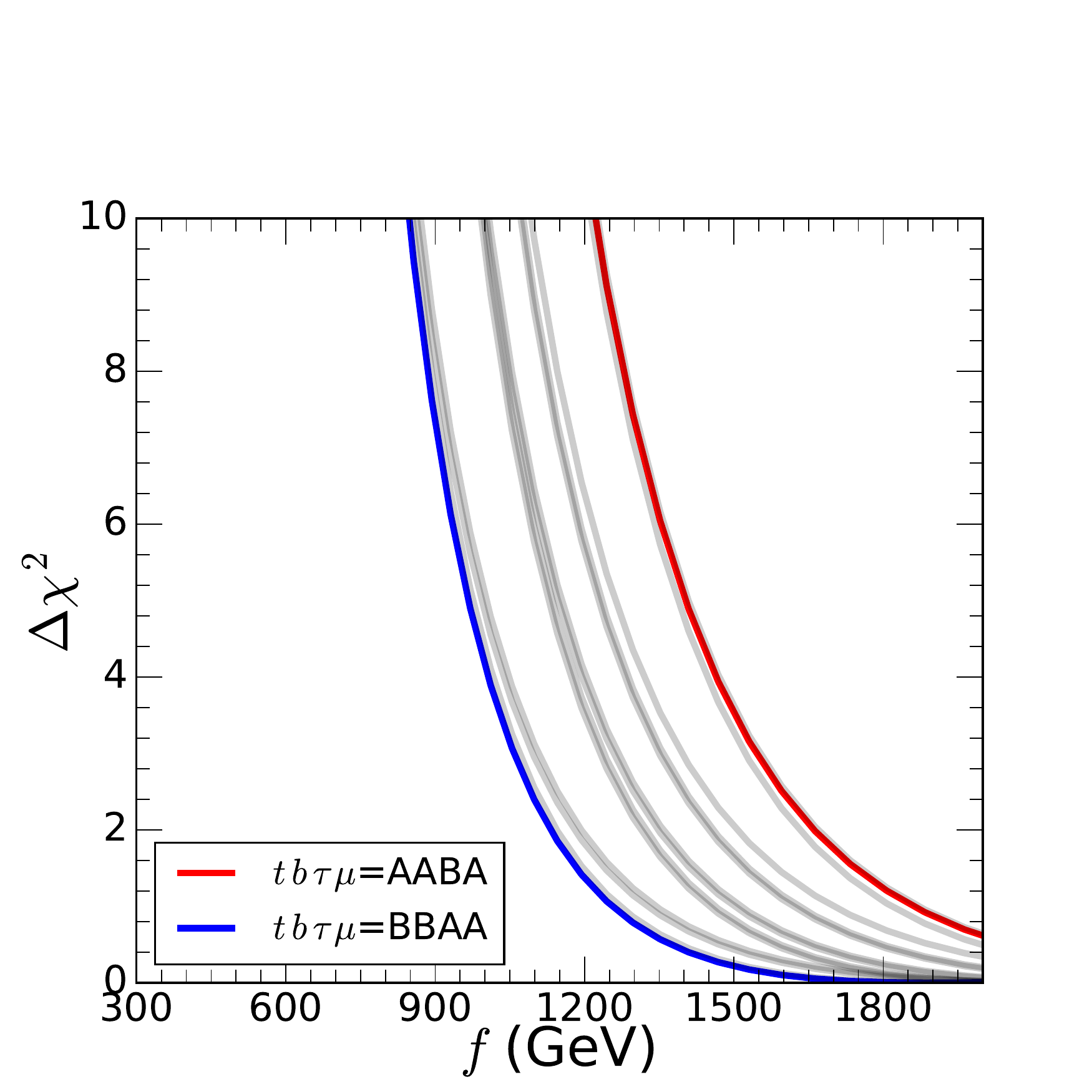}
\caption{$\Delta \chi^2$ as a function of the GB scale $f$ for several
  choices of the SM fermions embedding in minimal composite Higgs
  models.}
\label{fig:comph}
\end{figure}

Up to now we have focused on a bottom-up approach where all the Wilson
coefficients are treated as free parameters. Next, we consider the
minimal composite Higgs models and perform an analysis in which only
the deviations in Eqs.~\eqref{eq:amod} and~\eqref{eq:cmod}, parametrized
by the unique GB scale parameter $\xi=v^2/f$, are allowed.  We present
in Fig.~\ref{fig:comph} the $\Delta \chi^2$ dependence on the GB
scale, $f$, for several choices for the embedding of the SM top,
bottom, tau, and muon into the UV-model.  As we can see the least
stringent bound of $f$ originated for the choice $(B,B,A,A)$ for
$(t,b,\tau,\mu$) that reads $f > 1.$ TeV at 95\% CL, while the
strongest bound is for the choice $(A,A,B,A)$ for $(t,b,\tau,\mu$) and
implies $f > 1.45$ TeV. These results are in qualitative agreement
with the bounds derived Ref.~\cite{Khosa:2021wsu} with an slight
improvement of about $\sim$ 10--20\% from the more up-to-date data
samples considered.  The results in the figure are shown for the
analysis performed at linear order in the coefficients, but the
analysis performed at quadratic order leads to very similar
results. In particular this family of models do not allow for the
realization of the degenerated solution with inverse sign of the
couplings because, by construction, the $HVV$ and $Hff$ couplings in
these models have the same sign than in the SM.  The constraints
obtained for the minimal composite Higgs models also illustrate how
specific UV-completions are subject to stronger bounds due to
relations between the Wilson coefficients.  For example, the
marginalized 2$\sigma$ allowed range in Table~\ref{tab:ranges} for the
top Yukawa implies an upper bound on $f> 460\; (800)$ for models
embedding A (B) which are about a factor $\sim 2$ weaker than the
bounds in Fig.~\ref{fig:comph}. \medskip

In summary, the increased integrated luminosity gathered at LHC run 2
allows for improved tests of electroweak symmetry breaking scenarios
with a dynamical Higgs. We find no indication of statistically
significant deviations from the SM predictions in the analysis of
Higgs results and EWDBD.  This, in combination with the EWPO, results
into tighter constraints on the nature of the Higgs boson.\medskip

\acknowledgments

M.M. thanks the hospitality of the Universitat de Barcelona where part
of this work was done.  This work is supported in part by Conselho
Nacional de Desenvolvimento Cent\'{\i}fico e Tecnol\'ogico (CNPq)
grant 305762/2019-2, and by Funda\c{c}\~ao de Amparo \`a Pesquisa do
Estado de S\~ao Paulo (FAPESP) grants 2018/16921-1 and 2021/08669-3.
M.C.G-G is supported by spanish grant PID2019-105614GB-C21 financed by
MCIN/AEI/10.13039/501100011033, by USA-NSF grant PHY-1915093, and by
AGAUR (Generalitat de Catalunya) grant 2017-SGR-929. The authors
acknowledge the support of European ITN grant
H2020-MSCA-ITN-2019//860881-HIDDeN.

\bibliography{references}
\end{document}